\documentclass{article}
\usepackage{graphicx}

\begin{document}
\title{Convergence of Implicit Difference Scheme for 1D Lagrangian Hydrodynamics coupled to Radiation Transport Equation}
\author{\bf{Karabi Ghosh and S.V.G.Menon}\\ \\ Theoretical Physics Division,\\ Bhabha Atomic Research Centre, Mumbai 400085}
\date{}
\maketitle

\begin{abstract}
A fully implicit finite difference scheme has been developed to 
solve the hydrodynamic equations coupled with radiation transport. 
Solution of the time dependent radiation transport equation is 
obtained using the discrete ordinates method and the energy flow 
into the Lagrangian meshes as a result of radiation interaction 
is fully accounted for. A tridiagonal matrix system is solved at 
each time step to determine the hydrodynamic variables implicitly. 
The results obtained from this fully implicit radiation 
hydrodynamics code in the planar geometry agrees well with the 
scaling law for radiation driven strong shock propagation in 
aluminium. For the point explosion problem the self similar 
solutions are compared with results for pure hydrodynamic case 
in spherical geometry and the effect of radiation energy transfer 
is determined. Having, thus, benchmarked the code, convergence of 
the method w.r.t. time step is studied in detail and compared with 
the results of commonly used semi-implicit method. It is shown that 
significant error reduction is feasible in the implicit method in 
comparison to the semi-implicit method, though at the cost of 
slightly more CPU time. 
\end{abstract}


\section{Introduction}
Radiation transport and its interaction with matter via emission, 
absorption and scattering of radiation have a substantial effect on 
both the state and the motion of materials in high temperature 
hydrodynamic flows occurring in  inertial confinement fusion (ICF), 
strong explosions and astrophysical systems \cite{Zeldovich}. For 
many applications the dynamics can be considered non-relativistic 
since the flow velocities are much less than the speed of light. In 
order to describe properly the dynamics of the radiating flow, it 
is necessary to solve the full time-dependent radiation transport 
equation as very short time scales 
($t_R\sim l/c\ \mbox{or}\  t_\lambda\sim \lambda_p/c$ corresponding 
to a photon flight time over a characteristic structural length $l$, 
or over a photon mean free path $\lambda_p$ ) are to be considered 
\cite{Mihalas}.Two methods commonly used are non-equilibrium 
diffusion theory \cite{Dai1}, \cite{Bates} and radiation heat 
conduction approximation \cite{Zeldovich}. The former is valid for 
optically thick bodies, where the density gradients are small and 
the angular distribution of photons is nearly isotropic. The 
conduction approximation is only applicable when matter and 
radiation are in local thermodynamic equilibrium, so that the 
radiant energy flux is proportional to temperature gradient, and 
for slower hydrodynamics time scales \cite{Zeldovich}. Use of 
Eddington's factor for closing the first two moment equations is 
yet another approach followed in radiation hydrodynamics 
\cite{Dai2}. Radiative phenomena occur on time scales that differ 
by many orders of magnitude from those characterizing hydrodynamic 
flow. This leads to significant computational challenges in the 
efficient modeling of radiation hydrodynamics.

In this paper we solve the equations of hydrodynamics and the time 
dependent radiation transport equation fully implicitly. The 
anisotropy in the angular distribution of photons is treated in a 
direct way using the discrete ordinates method. Finite difference 
analysis is used for the Lagrangian meshes to obtain the 
thermodynamic variables. The hydrodynamic evolution of the system 
is considered in a fully implicit manner by solving a tridiagonal 
system of equations to obtain the velocities. The pressures and 
temperatures are converged iteratively.

Earlier studies on the non-equilibrium radiation diffusion 
calculations show that the accuracy of the solution increases 
on converging the non-linearities within a time step and increasing 
benefit is obtained as the problem becomes more and more nonlinear 
and faster \cite{Knoll}, \cite{Ober}. In this work, by iteratively 
converging the thermodynamic variables, we observe a faster decrease 
in the $L_2$-Error as compared to the commonly used 
semi-implicit scheme.

The organization of the paper is as follows: In section \ref{model} 
we discuss the finite difference scheme for solving the hydrodynamic 
equations followed by the solution procedure of the radiation 
transport equation and their coupling together to result in an 
implicit radiation hydrodynamics code. Section \ref{results} 
presents the results obtained using this fully implicit 
one-dimensional radiation hydrodynamics code for the problems 
of shock propagation in aluminium and the point explosion problem. 
These benchmark results, thus, prove the validity of the methods. 
Next, extensive results of convergence studies w.r.t. time step are 
presented. These results using the full transport equation are new, 
though similar studies have been reported earlier within approximate 
methods \cite{Bates}, \cite{Knoll2}. Finally the conclusions of this 
paper are presented in section \ref{conc}.
\label{intro}

\section{Simulation model}
\label{model}
 
\subsection{Implicit finite difference scheme for solving the 
hydrodynamic equations using a Lagrangian grid.}
\label{finite}

\subsubsection{Grid structure}
\label{grid}
\begin{figure}
\begin{center}
\includegraphics{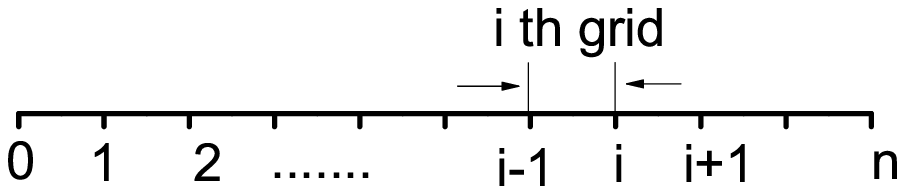}
\end{center}
\caption{\label{figure1}Grid structure.}
\end{figure}

For hydrodynamics calculations, the medium is divided into 
a number of cells as shown in Fig.  \ref{figure1}. The coordinate 
of the $i$ th vertex is denoted by $r_i$ and the region between 
the $(i-1)$ and $i$ th vertices is the $i$ th cell. The density 
of the $i$ th grid is $\rho_i$ and its mass is given by
\begin{eqnarray}
m_i=\acute c  \times \rho_i \times (r_i^\delta - r_{i-1}^\delta)
\end{eqnarray}
with $\acute c = 1, \pi, (4/3)\times \pi$ and $\delta=1,2,3$ for 
planar, cylindrical and spherical geometries respectively. Velocity 
of the $i$ th vertex is denoted by $u_i$ and 
$P_i,V_i,T_{ion,i},T_{elec,i},E_{ion,i}$ and $ E_{elec,i}$ are 
the total pressure, specific volume, temperature of ions and 
electrons and the specific internal energy of ions and 
electrons in the $i$ th mesh respectively. 

\subsubsection{Lagrangian step}
\label{step}
During a time interval $\Delta t\ $the vertexes $r_i$ 
of the cells move as 
\begin{eqnarray}
\label{pos}\tilde r_i = r_i+u_i^\star \Delta t\\
\label{vel}u_i^\star=(1/2)(u_i+\tilde u_i)
\end{eqnarray}
where $u_i^\star$ is the average of velocity values at the 
beginning and end of the Lagrangian step, $u_i$ and 
$\tilde u_i$ , respectively.

\subsubsection{Discretized form of the hydrodynamic equations}
\label{discretized}
In the Lagrangian formulation of hydrodynamics, the mass of each 
cell remains constant thereby enforcing mass conservation.

The Lagrangian differential equation for the conservation of 
momentum is 
\begin{eqnarray}
\label{momen}\rho \frac{d\vec u}{dt}=-\vec \nabla P
\end{eqnarray}
 Here, the total pressure is the sum of the electron, ion and 
radiation pressures i.e. $ P=P_{ion}+P_{elec}+P_{rad}$. 
Eq. [\ref{momen}] can be discretized for the velocity $\tilde u_i$ 
at the end of the time step in terms of the pressures,  
$P_i^{1/2} \mbox{and}\ P_{i+1}^{1/2}$ in the $i$ th and $i+1$ th 
meshes respectively, after half time step as \cite{Niem}
\begin{eqnarray}
\label{momentum}\tilde u_i = u_i - \frac{(P_{i+1}^{1/2}-P_i^{1/2})
\Delta t}{\rho_{i+1}(r_{i+1/2}-r_i)+\rho_i(r_i-r_{i-1/2})}
\end{eqnarray}

The velocity in the $i$ th mesh $\tilde u_i$ is determined by 
the pressure in the $i$ th and $i+1$ th meshes and hence all the 
meshes are connected. Mass conservation equation can be used to 
eliminate the pressures at half time step to obtain an equation 
relating the present time step velocities in the adjacent meshes 
as follows:

The equation describing conservation of mass is
\begin{eqnarray}
\label{mass}\frac{d\rho}{dt}=-\rho(\vec \nabla . \vec u)
\end{eqnarray}
where $\rho$ is the mass density of the medium. This equation can 
be rewritten in terms of pressure using the relation, 
$ \frac{dP}{dt}=(\frac{dP}{d\rho})_{S} \frac{d \rho}
{dt}=v^2 \frac{d\rho}{dt}$ where $v=\sqrt {(\frac{dP}{d\rho})_S}$
 is the
 adiabatic sound speed.
Therefore, Eq. [\ref{mass}] becomes
\begin{eqnarray}
\frac{dP}{dt}=-v^2 \rho \vec \nabla . \vec u
\end{eqnarray}
This can be written for all the one dimensional co-ordinate systems 
as
\begin{eqnarray}
\frac{dP}{dt}=-v^2 \rho \frac{1}{r^\alpha}\frac{d}{dr}r^\alpha u 
\end{eqnarray}
where $ \alpha = 0,1,2 $ for planar, cylindrical and spherical 
geometries.
This equation can be discretized to obtain the change in total 
pressure along a Lagrangian trajectory in terms of the velocity 
$\tilde u_i$ at the end of the time step \cite{Niem}:
\begin{eqnarray}
\label{pressure1}P_{i}^{1/2}=P_{i}+q_{i}- 
\rho_{i} v_{i}^2 \frac{1}{r_{i-1/2}^\alpha}
\times[\frac{r_{i}^\alpha \tilde u_{i} - r_{i-1}^ 
\alpha \tilde u_{i-1}}{r_{i}-r_{i-1}}]\frac{\Delta t}{2}
\end{eqnarray}
and
\begin{eqnarray}
\label{pressure2}P_{i+1}^{1/2}=P_{i+1}+q_{i+1}- 
\rho_{i+1} v_{i+1}^2 \frac{1}{r_{i+1/2}^\alpha}
\times[\frac{r_{i+1}^\alpha \tilde u_{i+1} - r_i^ 
\alpha \tilde u_i}{r_{i+1}-r_i}]\frac{\Delta t}{2}
\end{eqnarray}
Here, $q_i$ is the quadratic Von Neumann and Richtmyer 
artificial viscosity in the $i$ th mesh \cite{Von}.
\begin{eqnarray}
q_i=\frac{\acute k (\rho_i \Delta x_i)^2}{V_i}(\frac{dV_i}{dt})^2
\end{eqnarray}
where $\acute k $ ($\simeq 3$) is a dimensionless constant.

Using Eqs. [\ref{pressure1}] and [\ref{pressure2}] , 
$P_i^{1/2} \mbox{and}\ P_{i+1}^{1/2}$ in Eq.[\ref{momentum}] 
are eliminated to obtain a tridiagonal system of equations 
for $\tilde u_i$.
\begin{eqnarray}
-A_i \tilde u_{i+1}+B_i \tilde u_i-C_i \tilde u_{i-1}=D_i\\
\end{eqnarray}
where
\begin{eqnarray}
A_i=\frac{\rho_{i+1}(v_{i+1}\Delta t)^2}{2(\rho \Delta r)_i}
\times \frac{r_{i+1}^\alpha}{r_{i+1/2}^\alpha(r_{i+1}-r_i)}\\
B_i=1+\frac{\rho_{i+1}(v_{i+1}\Delta t)^2}{2(\rho \Delta r)_i}
\times \frac{r_i^\alpha}{r_{i+1/2}^\alpha(r_{i+1}-r_i)}\nonumber\\
+\frac{\rho_i(v_i\Delta t)^2}{2(\rho \Delta r)_i}\times 
\frac{r_i^\alpha}{r_{i-1/2}^\alpha(r_i-r_{i-1})}\\
C_i=\frac{\rho_i(v_i\Delta t)^2}{2(\rho \Delta r)_i}
\times \frac{r_{i-1}^\alpha}{r_{i-1/2}^\alpha(r_i-r_{i-1})}\\
\label{di}D_i=u_i-\frac{\Delta t}{(\rho \Delta r)_i}
[P_{i+1}+q_{i+1}-P_i-q_i]\\
\mbox{with}\nonumber\\
(\rho \Delta r)_i=\rho_{i+1}(r_{i+1/2}-r_i)+\rho_i(r_i-r_{i-1/2})
\end{eqnarray}
The energy equations, for the ions and electrons, 
expressed in terms of temperature are
\begin{eqnarray}
\rho[C_{Vion} \frac{\partial T_{ion}}{\partial t}+
\frac{\partial E_{ion}}{\partial V}\frac{\partial V}
{\partial t}]=-\frac{P_{ion}}{V}\frac{\partial V}{\partial t}-P_{ie}
\end{eqnarray}
and
\begin{eqnarray}
\rho[C_{Velec} \frac{\partial T_{elec}}{\partial t}+
\frac{\partial E_{elec}}{\partial V}\frac{\partial V}{\partial t}]
=-\frac{(P_{elec}+P_{rad})}{V}\frac{\partial V}{\partial t}+\nonumber\\
\sigma_R(T_{elec})[E_R(r,T_{elec})-B(T_{elec})]+P_{ie}
\end{eqnarray}
where $E_{ion} \ \mbox{and}\ E_{elec}$ are the specific internal 
energies and $V$ is specific volume. $\sigma_R(T_{elec})$ is the 
Rosseland opacity, $E_R(r,T_{elec})$ is the radiation energy flux 
and $\sigma_R(T_{elec})\ B(T_{elec})$ is radiation emission rate. 
$P_{ie}$ is the ion-electron energy exchange term given by
\begin{eqnarray}
P_{ie}(Tergs/cm^3/\mu s)=2.704 \times 10^{-40} n_{elec}\ 
n_{ion}\nonumber\\ 
\times \frac{T_{ion}-T_{elec}}{T_{elec}^{1.5}} M^{-1}Z^2 
\times \ln \Lambda 
\end{eqnarray}
with ion and electron temperatures expressed in keV. Further, 
'$n_{elec}$' and '$n_{ion}$' are the number densities of electrons 
and ions, M is the mass number and Z is the charge of the ions. 
Here the Coulomb logarithm for ion-electron collision
is \cite{Huba} 
\begin{eqnarray}
\ln \Lambda =\mbox{max}\{ 1, \ (23 - \ln[(n_{elec})^{0.5} Z\ 
T_{elec}^{-1.5}]) \}
\end{eqnarray}
with $T_{elec}$ expressed in eV. 

The discrete form of the energy equations for ions and electrons are

\begin{eqnarray}
T_{ion,i}^{n,k}=T_{ion,i}^{n-1}-(P_{ion,i}^{n,k-1}\ \Delta V_i^{n,k}+
\frac{P_{ie}^{n,k-1}\Delta t}{\rho_i^{n,k-1}}+\delta_{ion}^{n,k-1}
\Delta V_i^{n,k})/C_{Vion,i}^{n,k-1}
\end{eqnarray}
and
\begin{eqnarray}
\label{elece}T_{elec,i}^{n,k}=T_{elec,i}^{n,k-1}+
\frac{\rho_i^{n,k-1} C_{Velec,i}^{n,k-1}
(T_{elec,i}^{n-1}-T_{elec,i}^{n,k-1})}{\Delta t D_i^{n,k-1}}
\nonumber\\
+(\frac{\sigma_{Ri}^{n,k-1}}{D_i^{n,k-1}})(E_i^{n,k}-B_i^{n,k-1})
-\frac{(P_{elec}^{n,k-1}+P_{rad}^{n,k-1}+\delta_{elec}^{n,k-1})
\rho_i^{n,k-1}\Delta V_i^{n,k}}{\Delta t D_i^{n,k-1}}\nonumber\\
+P_{ie}^{n,k-1}/D_i^{n,k-1}
\end{eqnarray}
where
\begin{eqnarray}
 D_i^{n,k-1}=\frac{\rho_i^{n,k} C_{Velec,i}^{n,k-1}}
{\Delta t_n}+\sigma_{Ri}^{n,k-1}C_{v,Ri}^{n,k-1}\\
\delta_{ion}^{n,k-1}=(\frac{\partial  E_{ion}}
{\partial  V})_i^{n,k-1}\\
\delta_{elec}^{n,k-1}=(\frac{\partial  E_{elec}}
{\partial  V})_i^{n,k-1}\\
C_{vRi}^{n,k-1}=4ac(T_{elec,i}^{n,k-1})^3
\end{eqnarray}
with 'n' and 'k' denoting the time step and iteration 
index respectively. Also, the constants a 
(= $4\sigma_R/c$), $\sigma_R$ and c denote the radiation constant, 
Rosseland opacity and the speed of light respectively. 
Stefan-Boltzmann law, $B(T_{elec})= acT_{elec}^4 $, 
has been used explicitly in these equations.

\subsection{Discrete ordinates method for solving 
the radiation transport equation.}
\label{transport}
In the Gray approximation, or one group model, the 
time dependent radiation transport equation in a 
stationary medium is 
\begin{eqnarray}
\frac{1}{c}\frac{\partial I}{\partial t}+
\vec \Omega .\vec \nabla I+(\sigma _R(T)+\sigma_s )
I(\vec r,\vec \Omega,t )=\sigma _R(T)B(T)\nonumber\\
+\frac{\sigma_s}{4\pi}\int I(\vec r,\vec \Omega\acute  ,t )
d\vec\Omega\acute. 
\end{eqnarray}
where $I(\vec r,\vec \Omega,t)$ is the radiation energy flux, 
due to photons moving in the direction $\vec \Omega$, at space 
point $\vec r$ and time t. Here $\sigma_R(T)$ is the one group 
radiation opacity, which is assumed to be calculated by Rosseland 
weighing, at electron temperature $T$ (the subscript of $T_{elec}$ 
is dropped for convenience). As already mentioned, $B(T)$ is the 
radiation energy flux emitted by the medium which is given by the 
Stefan-Boltzmann law $B(T)=acT^4$. The radiation constant $a$ is 
$\simeq 137$ if $T$ is in $keV$ and $c$ in $cm/\mu s $. This 
formula for the emission rate follows from the local thermodynamic 
equilibrium (LTE) approximation, which is assumed in the present 
model. The scattering cross-section $\sigma_s$, representing 
Thomson scattering is assumed to be isotropic and independent 
of temperature. In the Lagrangian framework the radiation 
transport equation for a planar medium is 
\begin{eqnarray}
\frac{1}{c}\rho \frac{\partial}{\partial t}(\frac{I}{\rho})+
\mu \frac{\partial I}{\partial x}+(\sigma _R(T)+\sigma_s )
I(x,\mu,t)=\sigma_R(T)B(T)\nonumber\\+\frac{\sigma_s}{2}
\int_{-1}^{1} I(x,\mu\prime,t )d\mu\prime. 
\end{eqnarray} 
where $I(x,\mu,t)$ is the radiation energy flux along a 
direction at an angle $cos^{-1}(\mu)$ to the x axis. The term 
$\rho \frac{\partial}{\partial t}(\frac{I}{\rho})$ in this equation 
arises due to the Lagrange scheme used in solving the hydrodynamic 
equations.

Backward difference formula for the time derivative gives
\begin{eqnarray}
\label{back}\mu \frac{\partial I^{n,k}}{\partial x}+[\sigma_R^{n,k-1}
+(c\Delta t)^{-1}+\sigma_s]I^{n,k}=\sigma_R^{n,k-1}B^{n,k-1}
\nonumber\\+\frac{\sigma_s}{2}\int_{-1}^{1} I^{n,k}(\mu\prime )
d\mu\prime +\frac{\rho^{n,k-1}}{\rho^{n-1}}I^{n-1}(c\Delta t)^{-1}
\end{eqnarray}
Here, 'n' and 'k' denote the time step and iteration index for 
temperature respectively as earlier. This iteration arises because 
the opacity $\sigma_R(T)$ and the radiation emission rate 
$\sigma_R(T)B(T)$ are functions of the local temperature $T$. The 
converged spatial temperature distribution is assumed to be known 
for the hydrodynamic cycle for the previous time step. Starting 
with the corresponding values of $\sigma_R(T)$ and $B(T)$ , denoted 
by $\sigma_R^{n,0}$ and $B^{n,0}$, the radiation energy fluxes are 
obtained from the solution of the transport equation 
Eq. [\ref{back}]. The method of solution, well known in neutron 
transport theory, is briefly discussed below. This is used in the 
electron energy equation of hydrodynamics Eq. [\ref{elece}] to 
obtain a new temperature distribution and corresponding values of 
$\sigma_R^{n,1}$ and $B^{n,1}$. The transport equation is again 
solved using these new estimates and the iterations are continued 
until the temperature distribution converges.

Finally the transport equation can be expressed in conservation 
form in spherical geometry as
\begin{eqnarray}
\label{sph}\frac{\mu}{r^2} \frac{\partial}{\partial r}
(r^2 I^{n,k})+\frac{\partial}{\partial \mu}
[\frac{(1-\mu^2)I^{n,k}}{r}]+\sigma I^{n,k} =Q(r,\mu)
\end{eqnarray}
with
\begin{eqnarray}
\sigma=\sigma_R^{n,k-1}+(c\Delta t)^{-1}+\sigma_s\\
Q(r,\mu)=\sigma_R^{n,k-1}B^{n,k-1}+\frac{\sigma_s}{2}
\int_{-1}^{1} I^{n,k}(\mu\prime )d\mu\prime +
\frac{\rho^{n,k-1}}{\rho^{n-1}}I^{n-1}(c\Delta t)^{-1}
\end{eqnarray}
where, the second term in Eq. [\ref{sph}] accounts for 
angular redistribution of photons during free flight. 
This term arises as a result of the local coordinate 
system used to describe the direction of propagation of photons. 
If this term is omitted, Eq. [\ref{sph}] reduces to that for 
planar medium and therefore a common method of solution can be 
applied.

In the semi-implicit method to be discussed later, the transport 
equation is solved only once per time step. Then, a slightly more 
accurate linearization \cite{Larsen} can be introduced in Eqs. 
[\ref{back}] and [\ref{sph}] by replacing $B^{n,0}=B^{n-1}$ 
with $B^{n,1}=B^n$. A first order Taylor expansion yields 
$ B^n=B^{n-1}+(\partial B/\partial T)^n(T^{n}-T^{n-1})$ 
from which $(T^{n}-T^{n-1})$ can be eliminated using 
Eq. [\ref{elece}]. However this modification is not necessary 
in the implicit method as the iterations are performed for 
converging the temperature distribution.

To solve Eq. [\ref{sph}], it is written in the discrete angle 
variable as \cite{Lewis}
\begin{eqnarray}
\label{dis}\frac{\mu _m}{r^2}\frac{\partial}{\partial r}(r^2 I_m)+
\frac{2}{r\omega _m}(\alpha _{m+1/2}I_{m+1/2}-
\alpha _{m-1/2}I_{m-1/2})+\sigma I_m=Q_m
\end{eqnarray}
where the indices 'n','k' on I have been supressed. Here m refers
 to a particular value of $\mu$ in the angular range [-1,1] which 
is divided into M directions. The parameter $\omega _m$ is the 
weight attached to this direction whose value has been fixed 
according to the Gauss quadrature and $\alpha _{m\pm 1/2}$ are 
the angular difference coefficients. $I_m$ and $I_{m\pm 1/2}$ 
are the fluxes at the centers and the edges of the angular 
cell respectively. The angle integrated balance equation 
for photons is satisfied if the "$\alpha$-coefficients" 
obey the condition
\begin{eqnarray}
\label{bal}\Sigma_{m=1}^{M}[\alpha _{m+1/2}I_{m+1/2}-
\alpha _{m-1/2}I_{m-1/2}]=0
\end{eqnarray}
As photons traversing along $\mu = \pm 1$ are not 
redistributed during the flight, the $\alpha$-coefficients 
also obey the boundary conditions
\begin{eqnarray}
\alpha_{1/2}=\alpha_{M+1/2}=0
\end{eqnarray}
For a spatially uniform and isotropic angular flux, Eq. [\ref{dis}] 
yields the recursion relation
\begin{eqnarray}
\alpha_{m+1/2}=\alpha_{m-1/2}-\omega_m \mu_m
\end{eqnarray}
as the flux $I(r,\mu)$ is a constant in this case.

The finite difference version of Eq. [\ref{dis}] in space is 
derived by integrating over a cell of volume $V_i$ bounded by 
surfaces $A_{i\pm 1/2}$ where $V_i=4 \pi \int_{r-1/2}^{r+1/2} 
r^2 dr=\frac{4\pi}{3}(r_{i+1/2}^3-r_{i-1/2}^3)$ and 
$A_{i\pm 1/2}=4\pi r_{i\pm 1/2}^2$. 
The discrete form of the transport equation in space and angle 
is thus obtained as
\begin{eqnarray}
\label{A}\frac{\mu_m}{V_i}[A_{i+1/2}I_{m,i+1/2}-A_{i-1/2}
I_{m,i-1/2}]+\frac{2(A_{i+1/2}-A_{i-1/2})}{\omega_m V_i}\nonumber\\
\times [\alpha_{m+1/2}I_{m+1/2,i}-\alpha_{m-1/2}I_{m-1/2,i}]+\sigma 
I_{m,i}=Q_{m,i}
\end{eqnarray}
The cell average flux and source are given by 
\begin{eqnarray}
I_{m,i}=\frac{1}{V_i}4\pi \int_{r_{i-1/2}}^{r_{i+1/2}} r^2 I_m(r)dr
\end{eqnarray}
and 
\begin{eqnarray}
Q_{m,i}=\frac{1}{V_i}4\pi \int_{r_{i-1/2}}^{r_{i+1/2}} r^2 Q_m(r)dr
\end{eqnarray}
respectively, where '$i$' specifies the spatial mesh.
As mentioned earlier, planar geometry equations are obtained if the 
terms involving $\alpha_{m\pm 1/2}$ are omitted and the replacements 
$V_i=r_{i+1/2}-r_{i-1/2}$ and $A_{i+1/2}=1$ are made. Thus, both 
geometries can be treated on the same lines using this approach.
The difference scheme is completed by assuming that the flux varies 
exponentially between the two adjacent faces of a cell both 
spatially and angularly so that the centered flux $I_{m,i}$ can 
be expressed as \cite{Bar}:
\begin{eqnarray}
I_{m,i}=I_{m,i-1/2}\ exp\ [-\frac{1}{2}(r_{i+1/2}-r_{i-1/2})]\\
I_{m,i}=I_{m,i+1/2}\ exp\ [+\frac{1}{2}(r_{i+1/2}-r_{i-1/2})]
\end{eqnarray}
where the radii $r_{i+1/2}$ and $r_{i-1/2}$ are expressed in 
particle mean free paths. These relations show that
\begin{eqnarray}
\label{B}I_{m,i}^2=I_{m,i-1/2}I_{m,i+1/2}
\end{eqnarray}
for the spatial direction. Similarly for the angular direction 
one gets
\begin{eqnarray}
\label{C}I_{m,i}^2=I_{m-1/2,i}I_{m+1/2,i}
\end{eqnarray}
Use of these difference schemes guarantees positivity of all 
the angular fluxes if {$Q_{m,i}$} are positive.
The symmetry of the flux at the centre of the sphere is 
enforced by the conditions
\begin{eqnarray}
\label{D}I_{M+1-m,1/2}=I_{m,1/2}, \ \ m=1,2,......,M/2
\end{eqnarray}
Dividing the spatial range into L intervals, for a vacuum 
boundary at $r_{L+1/2}$, we have 
\begin{eqnarray} 
\label{E}I_{m,L+1/2}=0 , \ \ m=1,2,.......,M/2
\end{eqnarray}
i.e, at the rightmost boundary the fluxes are zero for all 
directions pointing towards the medium. Alternately, boundary 
sources, if present, can also be specified.

An iterative method is used to solve the transport equation 
to treat the scattering term. The radiation densities at the 
centre of the meshes are taken from the previous time step, 
thereby providing the source explicitly. The fluxes $I
_{1/2,i}$ for all meshes do not occur in Eq. [\ref{A}] as 
$\alpha_{1/2}=0$. Then the fluxes $I_{3/2,i}$ are eliminated 
from this equation using the upwind scheme $I_{3/2,i}=I_{1,i}$. 
Starting from the boundary condition, viz. Eq. [\ref{E}], Eq. 
[\ref{A}] and [\ref{B}] can be used to determine these two fluxes 
for all the spatial meshes '$i$'. Thereafter together with Eq. 
[\ref{C}], the fluxes for all the negative values of $\mu_m$ can 
be solved for. At the center, the reflecting boundary condition 
given by Eq. [\ref{D}] provide the starting fluxes for the outward 
sweeps through all the spatial and angular meshes with positive 
values of $\mu_m$.

This completes one space-angle sweep providing new estimates of 
radiation energy flux at the mesh centers, given by:
\begin{eqnarray}
E_{Ri}^{n,k}=\sum_{m}\omega_m I_{m,i}/\sum_{m}\omega_m
\end{eqnarray}
where the sum extends over all directions M. The mesh-angle 
sweeps are repeated until the scattering source distribution 
converges to a specified accuracy. The rate of radiation energy 
absorbed by unit mass of the material in the $i$ th mesh is
\begin{eqnarray}
\varepsilon _i=\sigma_{Ri}^{n,k-1}[E_{Ri}^{n,k}-B_i^{n,k-1}]/
\rho^{n,k}_i
\end{eqnarray}
which determines the coupling between radiation transport and 
hydrodynamics.

\subsection{Implicit radiation hydrodynamics solution method}
\label{solution}
The sample volume is divided into 'L' meshes of equal width. 
The initial position and velocity of all the vertices are 
defined according to the problem under consideration. Also 
the initial pressure, temperature and internal energy of 
all the meshes are entered as input.

For any time step, the temperature of the incident radiation 
is obtained by interpolating the data for the radiation temperature 
as a function of time (as in the case of shock propagation in 
aluminium sheet or an ICF pellet implosion in a hohlraum).
All the thermodynamic parameters for this time step are initialized 
using their corresponding values in the previous time step. 
It is important to 
note that the velocity $u_i$ in Eqs. [\ref{vel}] and [\ref{di}] and 
position $r_i$ in Eq. [\ref{pos}] are the old variables and remain 
constant unless the pressure and temperature iterations for this 
time step converge.

The temperature iterations begin by solving the radiation transport 
equation for all the meshes which gives the energy flowing from 
radiation to matter. 

The 1D Lagrangian step is a leapfrog scheme where new radial 
velocities $\tilde u_i$ arise due to acceleration by pressure 
gradient evaluated at half time step. This leads to a time 
implicit algorithm.
The first step in the pressure iteration starts by solving the 
tridiagonal system of equations for the velocity of all the 
vertices. The sound speed is obtained from the Equation of 
state (EOS) of that material. The new velocities and positions 
of all the vertices are obtained which are used to calculate 
the new density and change in volume of all the meshes. The 
total pressure is obtained by adding the Von Neumann and 
Richtmeyer artificial viscosity to the ion, electron and radiation 
pressures and solving the energy equations which takes into account 
both the energy flow from radiation and the work done by (or on) 
the meshes due to expansion (or contraction).
The energy equations for ions and electrons are solved using the 
corresponding material EOS which provides the pressure and the 
specific heat at constant volume of the material (both ions and 
electrons).
The hydrodynamic variables like the position, density, internal 
energy and velocity of all the meshes are updated. The convergence 
criterion for the total pressure is checked and if the relative 
error is greater than a fixed error criterion, the iteration for 
pressure continues, i.e, it goes back to solve the tridiagonal 
equations to obtain the velocities, positions, energies and so on. 
When the pressure converges according to the error criterion, 
the convergence for the electron temperature is checked in a 
similar manner. The maximum value of the error in electron 
temperature for all the meshes is noted and if this value 
exceeds the value acceptable by the error criterion, the 
temperature iterations continue, i.e, transport equation, 
tridiagonal system of equations for velocity, etc, are solved, 
until the error criterion is satisfied. Thus the method is 
fully implicit as the velocities of all the vertices are obtained 
by solving a set of simultaneous equations. Also both the 
temperature and pressure are converged simultaneously using 
the iterative method.
Once both the pressure and temperature converge, the position 
of the shock front is obtained by noting the pressure change 
and the new time step is estimated as follows:

The time step $\Delta t$ is chosen so as to satisfy the Courant 
condition which demands that it is less than the time for a sound 
signal with velocity $v$ to traverse the grid spacing
 $\Delta x$, $\frac{v \Delta t}{\Delta x} < C $ where the 
reduction factor C is referred to as the Courant number. 
The stability  analysis of Von Neumann introduces additional 
reduction in time step due to the material compressibility 
\cite{Wilkins}.  

\begin{figure}
\begin{center}
\includegraphics[width=15cm]{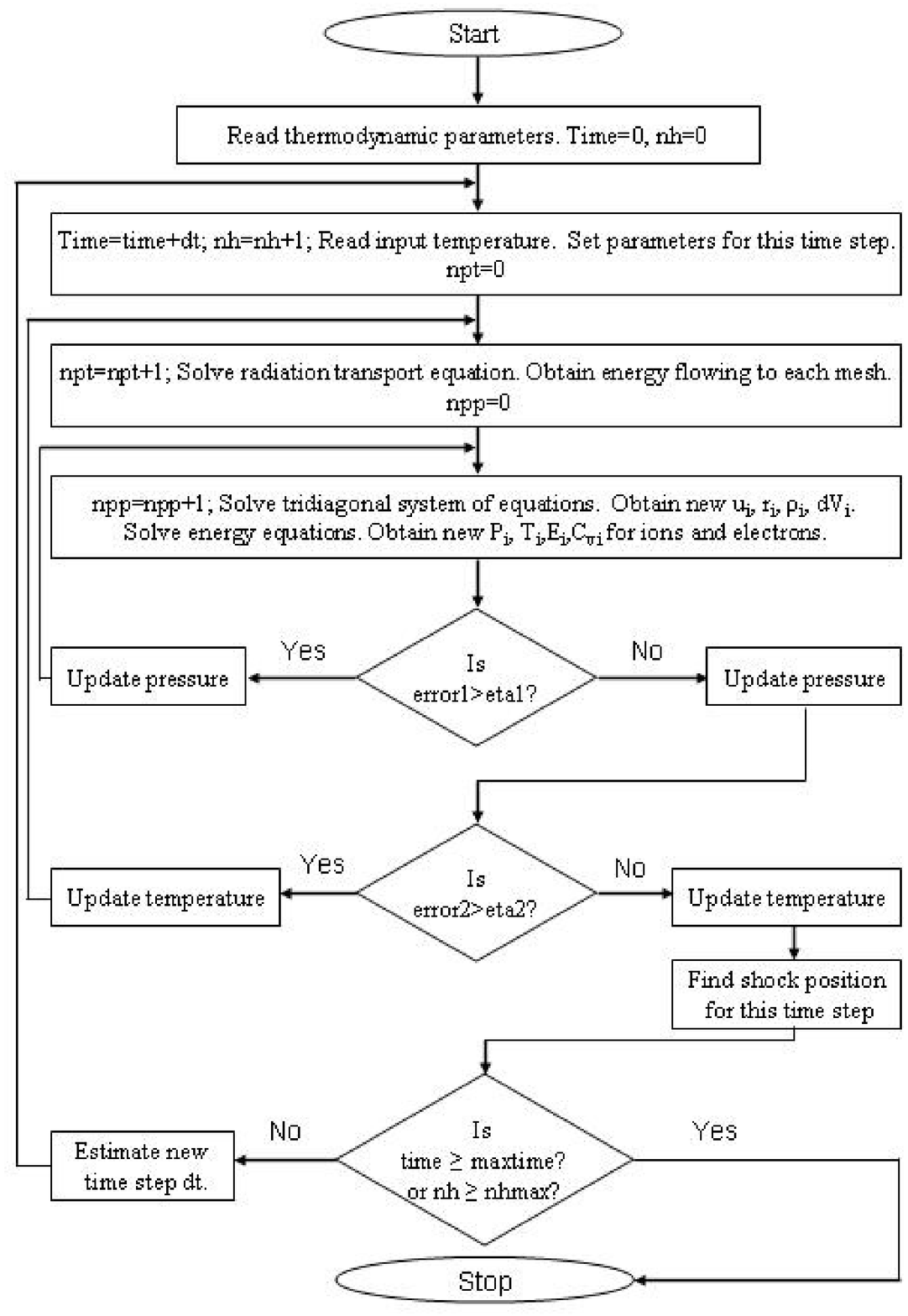}
\end{center}
\caption{\label{figure2}Flowchart for the Implicit 1D 
Radiation Hydrodynamics. Here, 'nh' is the time step index 
and 'dt' is the time step taken. The iteration indices for 
electron temperature and total pressure  are expressed as 
'npt' and 'npp' respectively. 'Error1' and 'Error2' are the 
fractional errors in pressure and temperature respectively 
whereas 'eta1' and 'eta2' are those acceptable by the error 
criterion.}
\end{figure}

The above procedure is repeated up to the time we are interested 
in following the evolution of the system.
The solution method described above is clearly depicted in the 
flowchart given in Fig. \ref{figure2}. The time step index is 
denoted by 'nh' and 'dt' is the time step taken. The iteration 
indices for electron temperature and total pressure  are 
expressed as 'npt' and 'npp' respectively. 'Error1' and 'Error2' 
are the fractional errors in pressure and temperature respectively 
whereas 'eta1' and 'eta2' are those acceptable by the error 
criterion.
\subsection{Semi-implicit method}\label{explicit}
In the semi-implicit scheme, Eq. [\ref{momentum}] is retained 
and $P_i^{1/2}$ is expressed as $P_i^{1/2} = (P_i+\tilde P_i)/2$ 
wherein $\tilde P_i$ is the pressure at the end of the time step. 
Starting with the previous time step values for $\tilde P_i$, the 
position and velocity of each mesh is obtained and $\tilde P_i$ 
is iteratively converged using the EOS. As the variables are 
obtained explicitly from the known values, there is no need to 
solve the tridiagonal system of equations for the velocities 
of all the meshes \cite{Richt}. Again, the energy flowing to 
the meshes as a result of radiation interaction is obtained 
by solving the transport equation once at the start of the 
time step, and hence the iterations leading to temperature 
convergence are absent. 
\section{Results}\label{results}
\subsection{Investigation of the performance of the scheme using 
benchmark problems}\label{benchmark}
\subsubsection{Shock propagation in Aluminium}\label{aluminium}

In the indirect drive inertial confinement fusion, high power 
laser beams are focused on the inner walls of high Z cavities 
or hohlraums, converting the driver energy to x-rays which implode
 the capsule. If the x-ray from the hohlraum is allowed to fall 
on an aluminium foil over a hole in the cavity, the low Z material
 absorbs the radiation and ablates generating a shock wave. Using 
strong shock wave theory, the radiation temperature in the cavity 
$T_R$ can be correlated to the shock velocity $u_s$. The scaling 
law derived for aluminium is $T_R=0.0126 u_s^{0.63}$, where $T_R$ 
is in units of eV and $u_s$ is in units of cm/s for a temperature 
range of 100-250 eV \cite{Kauffman}.

For the purpose of simulation, an aluminium foil of thickness 
0.6 mm and unit cross section is chosen. It is subdivided into 
300 meshes each of width $2 \times 10^{-4}$ cm. An initial guess 
value of $10^{-7}\mu s$ is used for the time step. The equilibrium 
density of Al is 2.71 gm/cc. In the discrete ordinates method four 
angles are chosen. As the temperature attained for this test problem 
is somewhat low, the total energy equation is solved assuming that 
electrons and ions are at the same temperature (the material 
temperature). The Equation of State (EOS) and Rosseland opacity 
for aluminium are given by 
\begin{eqnarray}
e = \frac{PV}{\gamma-1}=\epsilon T^\mu V^\nu
\end{eqnarray}

\begin{eqnarray}
\kappa_R = l^{-1} T^{-\mu_R} V^{-\nu_R} 
\end{eqnarray}
These power law functions, of temperature and density, where 
$\epsilon, l,\mu, \nu, \mu_R\ \mbox{and}\ \nu_R  $ are the 
fitting parameters, are quite accurate in the temperature 
range of interest \cite{Basko}.

\begin{figure}
\begin{center}
\includegraphics{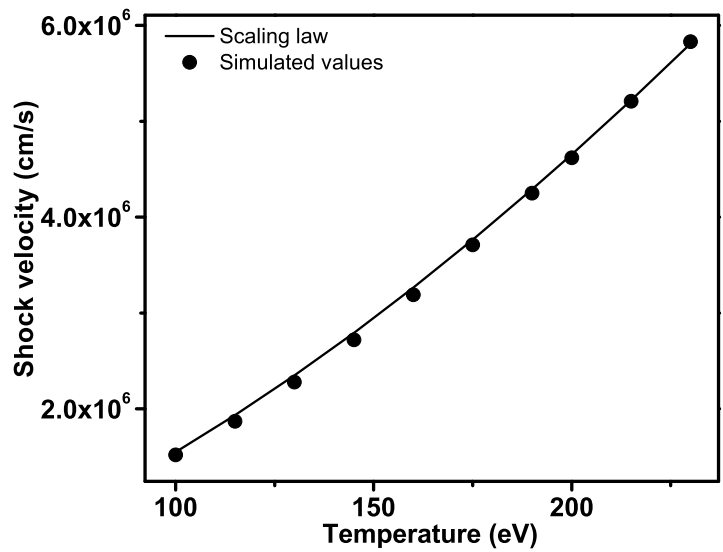}
\end{center}
\caption{\label{figure3}Comparison of simulation data 
(dots) with scaling law (line) relating shock velocity 
with the radiation temperature for aluminium.}
\end{figure}

Using the fully implicit radiation hydrodynamics code, a 
number of simulations are carried out for different values 
of time independent incident radiation fluxes or temperatures. 
Corresponding shock velocities are then determined after the 
decay of initial transients. In Fig. \ref{figure3}. we show the 
comparison between the numerically obtained shock velocities for 
different radiation temperatures (dots) and the scaling law for 
aluminium (line) mentioned earlier. Good agreement is observed 
in the temperature range where the scaling law is valid .

\begin{figure}
\includegraphics[width=7.5cm]{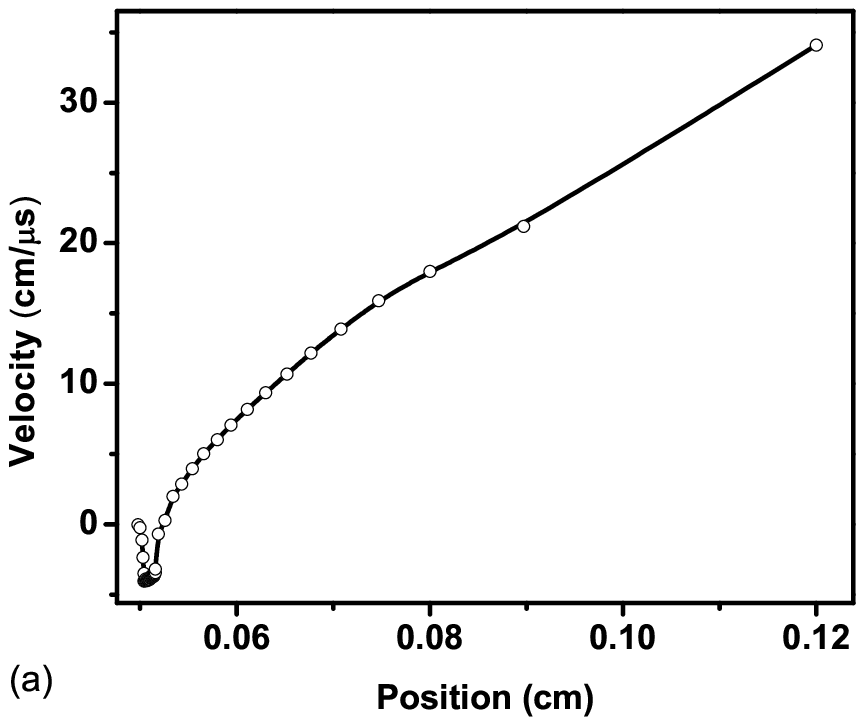}
\includegraphics[width=7.5cm]{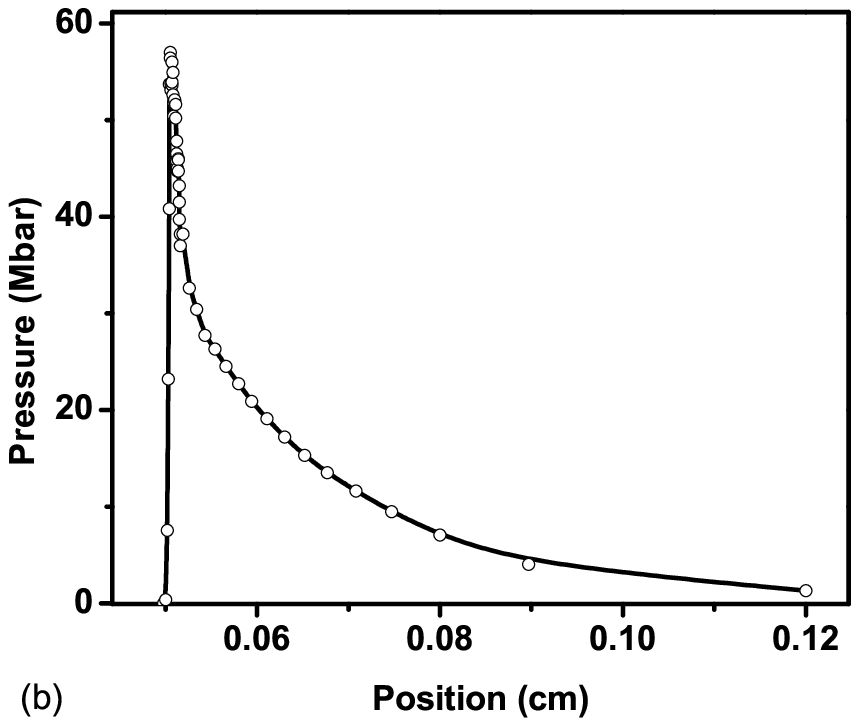}
\includegraphics[width=7.5cm]{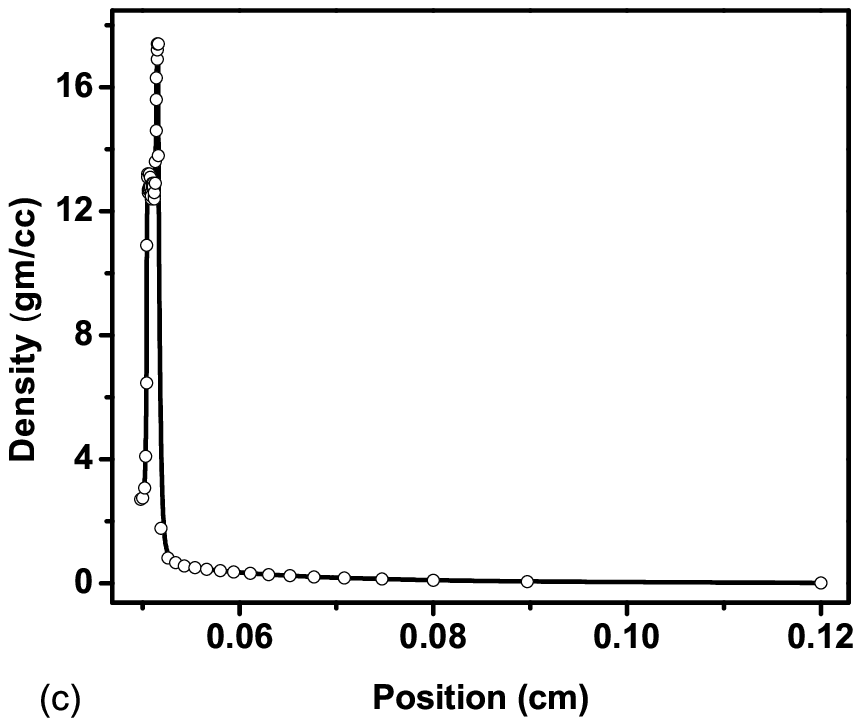}
\includegraphics[width=7.5cm]{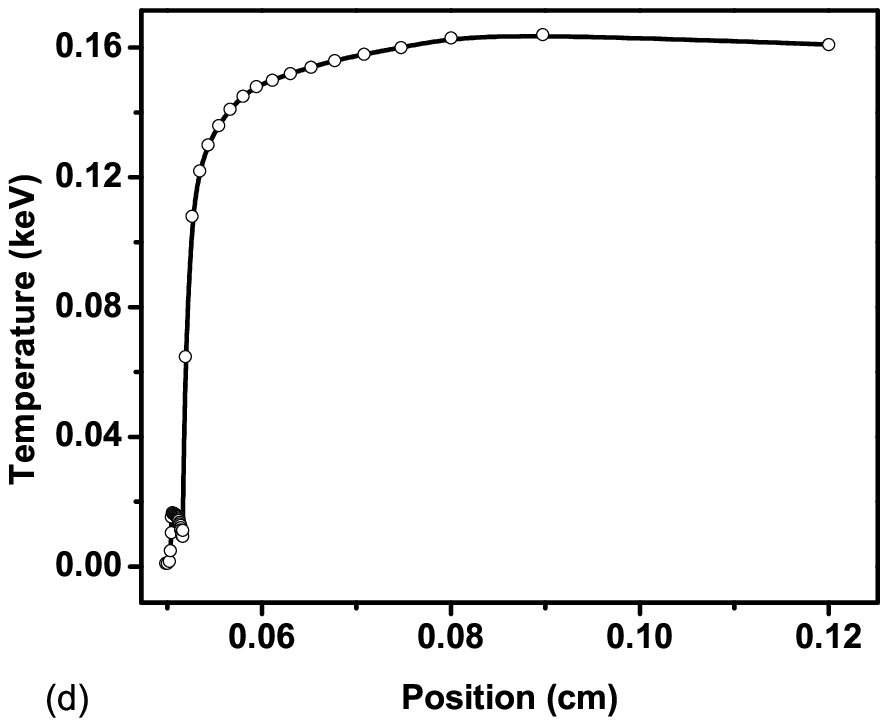}
\caption{\label{figure4}Profiles of the thermodynamic 
variables like (a) velocity, (b) pressure, (c) density 
and (d) temperature in the region behind the shock as a 
function of the distance at t = 2.5 ns. The region ahead 
of the shock is undisturbed and retain the initial values of 
the variables. The incident radiation temperature on the Al 
foil is shown in Figure 5.}
\end{figure}

\begin{figure}
\begin{center}
\includegraphics{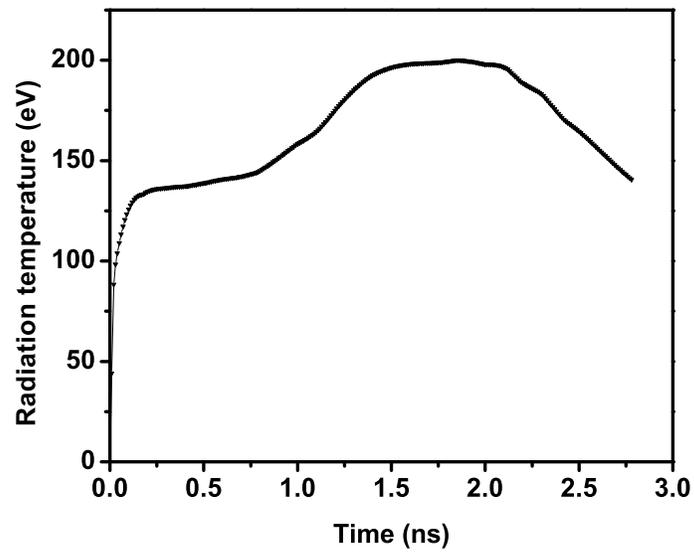}
\end{center}
\caption{\label{figure5}Radiation temperature profile in the 
hohlraum for strong shock propagation in aluminium.}
\end{figure}

\begin{figure}
\begin{center}
\includegraphics{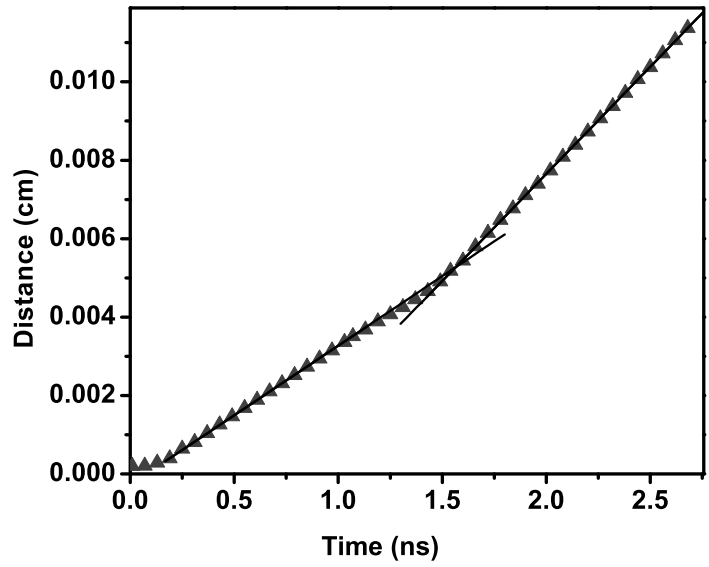}
\end{center}
\caption{\label{figure6}Distance traversed by the shock front 
vs. time graph in Al foil for the incident radiation temperature 
shown in Figure 5. The two slopes correspond to the two plateaus 
in the radiation profile.}
\end{figure}

Fig. \ref{figure4}. shows the various thermodynamic variables 
like velocity, pressure, density and  material temperature 
after 2.5 ns when the radiation profile shown in Fig. \ref{figure5}. 
is incident on the outermost mesh . This radiation temperature 
profile (Fig. \ref{figure5}. ) is chosen so as to achieve a 
nearly isentropic compression of the fuel pellet. The pulse 
is shaped in such a way that the pressure on the target surface 
gradually increases, so that the generated shock rises in 
strength. From Fig. \ref{figure4}. we observe that the outer 
meshes have ablated outwards while a shock wave has propagated 
inwards. At 2.5 ns, the shock is observed at 0.5 mm showing a peak 
in pressure and density. As the outer region has ablated, they 
move outwards with high velocities. The outermost mesh has moved 
to 1.2 mm. The meshes at the shock front move inwards showing 
negative velocities. Also the temperature profile shows that 
the region behind the the shock gets heated to about 160 eV. 
In Fig. \ref{figure6}. we plot the distance traversed by the 
shock front as a function of time for the above radiation 
temperature profile. The shock velocity changes from 3.54 to
 5.46 $cm/\mu s$ at 1.5 ns when the incident radiation 
temperature increases to 200 eV.  

The performance of the implicit and semi-implicit schemes are 
compared by studying the convergence properties and the CPU cost 
for the problem of shock wave propagation in aluminium. The 
convergence properties are examined by obtaining the absolute 
$L_2$-Error in the respective thermodynamic variable profile 
versus the fixed time step value. The absolute $L_2$-Error in 
the variable $f$ (velocity, pressure, density or temperature) 
is defined as
\begin{eqnarray}
L_2\mbox{-Error}=\label{converge}[\sum_{j=1}^{N}(f_j-f_j^e)^2]
^{1/2}
\end{eqnarray}
where the data $f_j^e$ constitute the exact solution for 
$\Delta t \longrightarrow 0$. This exact solution is chosen 
to be the result from a run using the implicit method with a 
small time step value of $10^{-10}\mu s$. The summation is taken 
over the values in all the meshes.

\begin{figure}
\includegraphics[width=7.5cm]{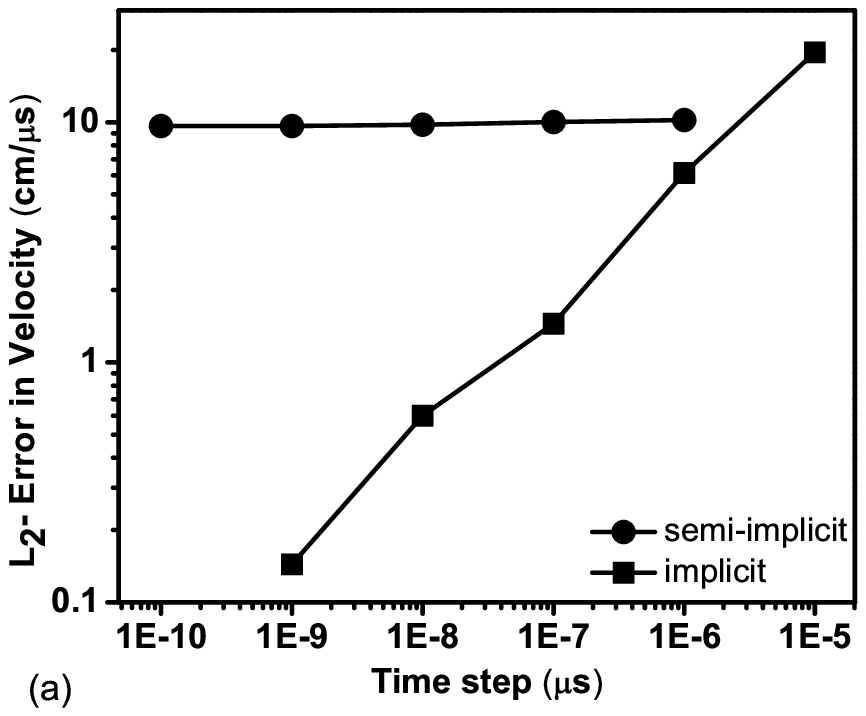}
\includegraphics[width=7.5cm]{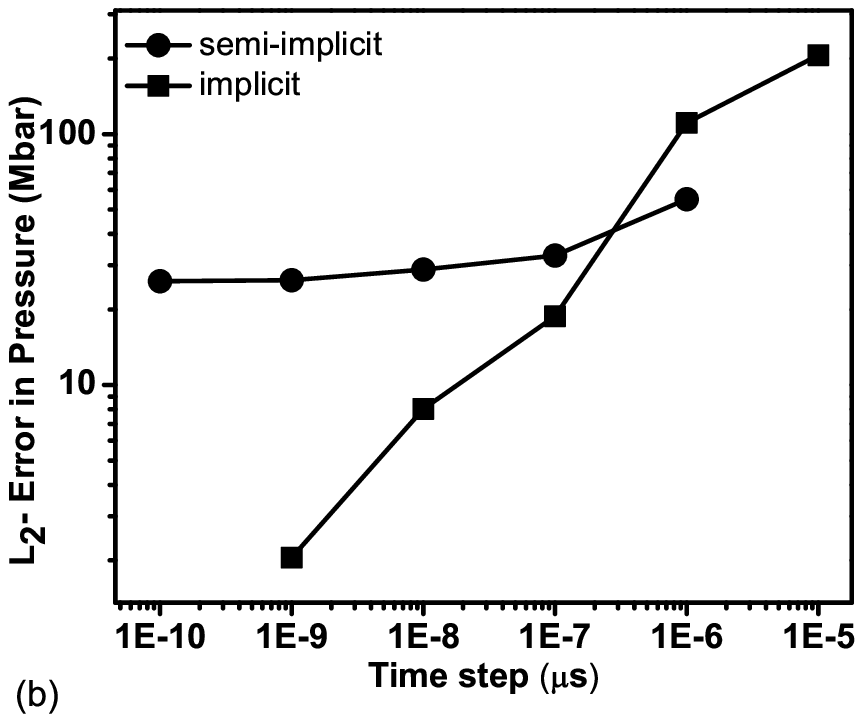}
\includegraphics[width=7.5cm]{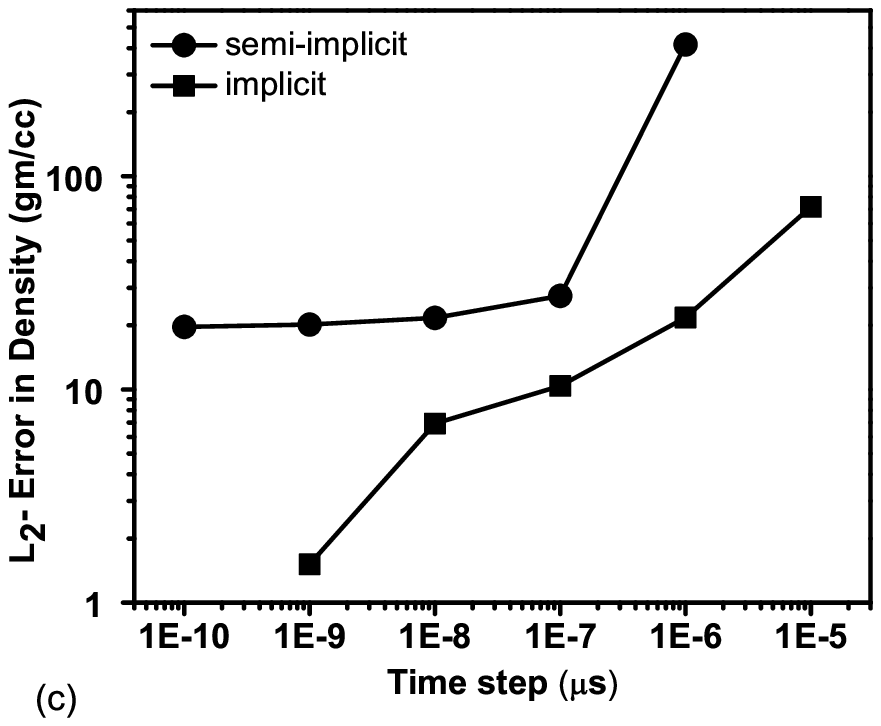}
\includegraphics[width=7.5cm]{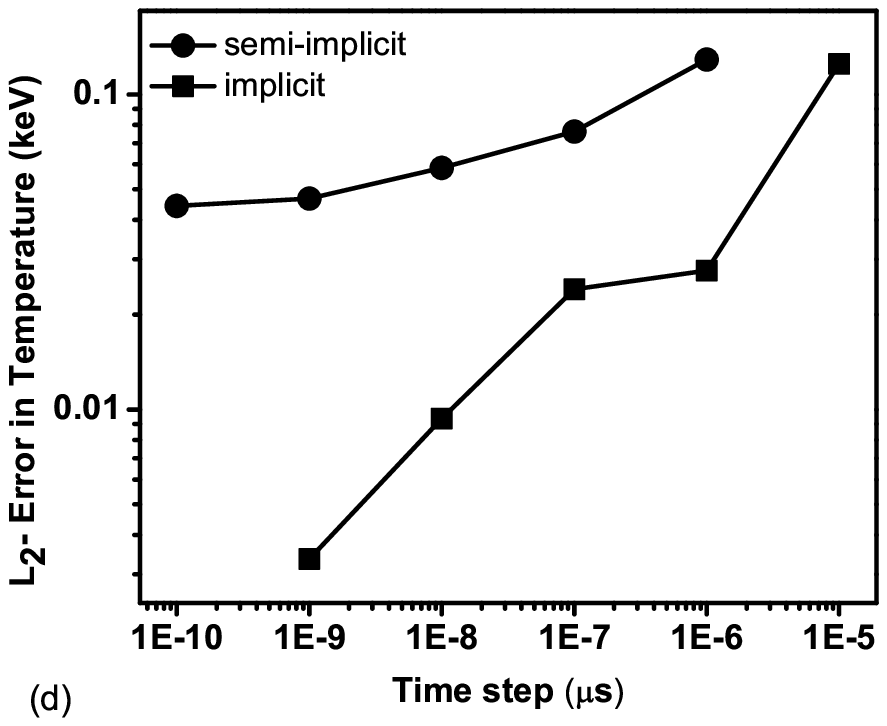}
\caption{\label{figure7}Comparison of the $L_2$-Error Vs. time step 
for the shock wave propagation problem in aluminium with mesh width 
$\Delta x = 2 \times 10^{-4}$ cm. Convergence is faster in the 
implicit scheme. }
\end{figure}

\begin{figure}
\begin{center}
\includegraphics{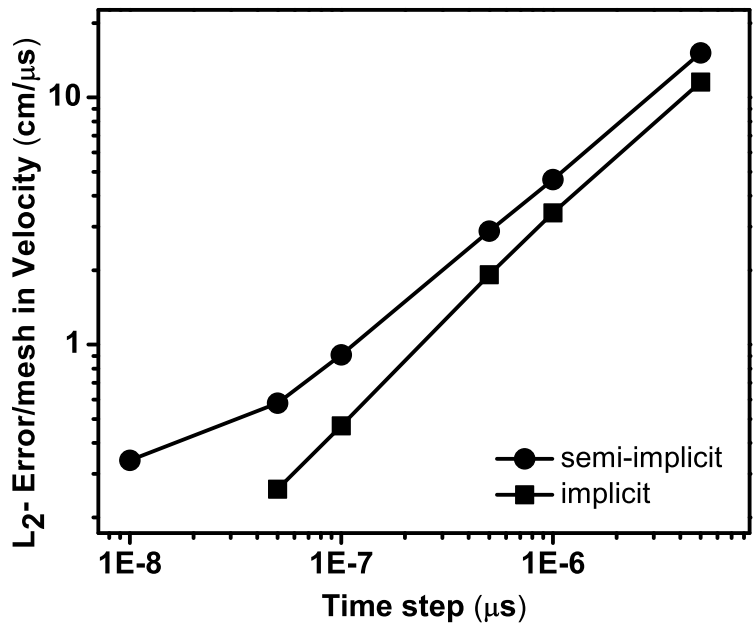}
\end{center}
\caption{\label{figure8}$L_2$-Error/mesh in velocity Vs. time step 
for the shock wave propagation problem in aluminium with 
$\Delta t/ \Delta x=5 \times 10^{-3} \mu s/ cm$. Convergence rate 
is higher for the implicit scheme.}
\end{figure}

Fig. \ref{figure7}. shows the absolute $L_2$-Error versus the 
time step value for velocity, pressure, density and temperature 
obtained using the implicit and semi-implicit radiation 
hydrodynamics codes respectively. The semi-implicit differencing 
scheme fails for time steps of $10^{-5}\mu s$ and higher because 
of the violation of CFL criterion. For time steps 
$\sim 10^{-6}\mu s$, the errors obtained from the two schemes 
are comparable. But as the time steps are reduced, the implicit 
scheme converges very fast, i.e. the errors reduce quickly, 
whereas the error becomes nearly constant in the semi-implicit 
scheme because of the spatial discretization error. An initial 
mesh width of $2 \times 10^{-4}$ cm is chosen for the above 
convergence study, which prevents further decrease in error 
in the semi-implicit scheme. Hence a reduction in the mesh 
width as well as the time steps is expected to decrease the error.
In Fig. \ref{figure8}. the $L_2$-Error per mesh for velocity i.e. 
$\label{converge}[\sum_{j=1}^{N}(f_j-f_j^e)^2/
\mbox{number of meshes}]^{1/2}$, is plotted as a function of the
 time step by keeping the ratio of time step to mesh size i.e. 
$\Delta t/ \Delta x$ constant at $5 \times 10^{-3} \mu s/ cm$. 
The results obtained from the implicit method using a small time
 step of $\Delta t=10^{-8} \mu s$ and mesh width of 
$2 \times 10^{-6}$ cm is chosen as the exact solution. 
Both the implicit and semi-implicit schemes show linear 
convergence, though the convergence rate is faster for 
the implicit scheme showing its superiority in obtaining 
higher accuracies. 

\begin{figure}
\begin{center}
\includegraphics{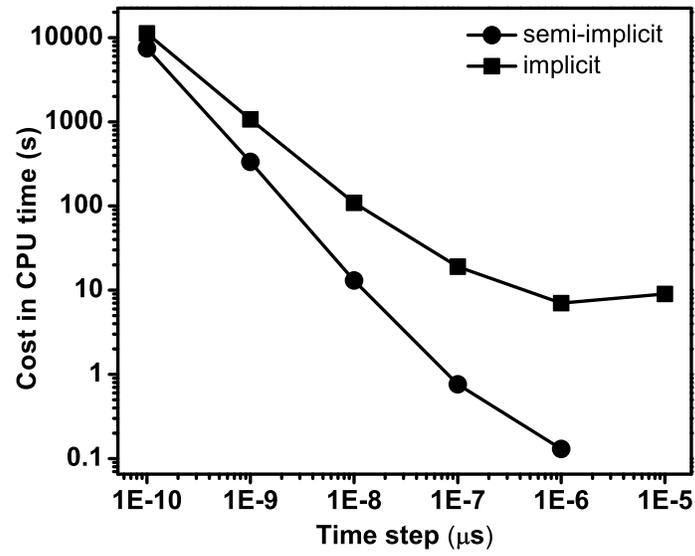}
\end{center}
\caption{\label{figure9}CPU cost Vs. time step for the shock wave 
propagation problem in aluminium with mesh width 
$\Delta x = 2 \times 10^{-4}$ cm.}
\end{figure}

Fig. \ref{figure9}. shows that the faster convergence in the 
implicit method (Fig. \ref{figure7}.) is attained at the cost 
of slightly higher CPU time. However the cost in CPU seconds 
become comparable in the two schemes for smaller time steps.
 All the runs in this study were done on a Pentium(4) computer
 having 1GB of RAM operating at 3.4 GHz.

\subsubsection{Point explosion problem}\label{point}
The self similar problem of a strong point explosion was formulated 
and solved by Sedov \cite{Sedov}. The problem considers a perfect 
gas with constant specific heats and density $\rho_0$ in which a 
large amount of energy $E$ is liberated at a point instantaneously.
 The shock wave propagates through the gas starting from the point 
where the energy is released. For numerical simulation, the energy
 E is assumed to be liberated in the first two meshes. The process 
is considered at a larger time $t$ when the radius of the shock 
front $R(t)>> r_0$, the radius of the region in which energy is 
released. It is also assumed that the process is sufficiently 
early so that the shock wave has not moved too far from the 
source. This ascertains that the shock strength is sufficiently 
large and it is possible to neglect the initial gas pressure $P_0$ 
or counter pressure in comparison with the pressure behind the 
shock wave \cite{Zeldovich}.

Under the above assumptions the gas motion is determined by four 
independent variables, viz, amount of energy released $E$, initial
 uniform density $\rho_0$, distance from the centre of the explosion 
$r$ and the time $t$.
The dimensionless quantity $\xi =r/R$ serves as the similarity 
variable. The motion of the wavefront $R(t)$ is governed by 
the relationship
\begin{eqnarray}
R=\xi_0(\frac{E}{\rho_0})^{1/5}t^{2/5} 
\end{eqnarray} 
where $\xi_0 $ is an independent variable.
The propagation velocity of the shock wave is 
\begin{eqnarray}
D=\frac{2}{5}\xi_0^{5/2}(\frac{E}{\rho_0})^{1/2}R^{-3/2}
\end{eqnarray}
The parameters behind the shock front using the limiting 
formulas for a strong shock wave are
\begin{eqnarray}
u_1=\frac{2}{\gamma+1}D\\
P_1=\frac{2}{\gamma+1}\rho_0 D^2\\
\rho_1=\rho_0 \frac{\gamma+1}{\gamma-1}\\
T_1=\frac{P_1}{(\gamma-1)\rho_1 C_V}
\end{eqnarray}
where $C_V$ is the specific heat at constant volume and 
$\gamma=C_P/C_V$ is the ratio of the specific heats. The 
distributions of velocity, pressure and density w.r.t. the 
radius are determined as functions of the dimensionless 
variable $\xi = r/R$. Since the motion is self-similar, 
the solution can be expressed in the form
\begin{eqnarray}
\label{soln}\ \ u=u_1(t)\tilde u(\xi),\ \ P=P_1(t)\tilde P 
(\xi),\ \ \rho=\rho_1 \tilde \rho(\xi)
\end{eqnarray}  
where $\tilde u, \tilde P\ \mbox{and}\  \tilde \rho$ are new 
dimensionless functions. The hydrodynamic equations, which are 
a system of three PDE's, are transformed into a system of three 
ordinary first-order differential equations for the three unknown 
functions $\tilde u, \tilde P\ \mbox{and}\  \tilde \rho$ by 
substituting the expressions given by Eq. [\ref{soln}] into 
the hydrodynamic equations for the spherically symmetric case 
and transforming from r and t to $\xi$. The boundary condition 
satisfied by the solution at the shock front ($r=R\ $  or $\xi=1$) 
is $\tilde u=\tilde P= \tilde \rho=1$. The dimensionless parameter 
$\xi_0$, which depends on the specific heat ratio $\gamma$ is 
obtained from the condition of conservation of energy evaluated 
with the solution obtained. 

\begin{figure}
\begin{center}
\includegraphics{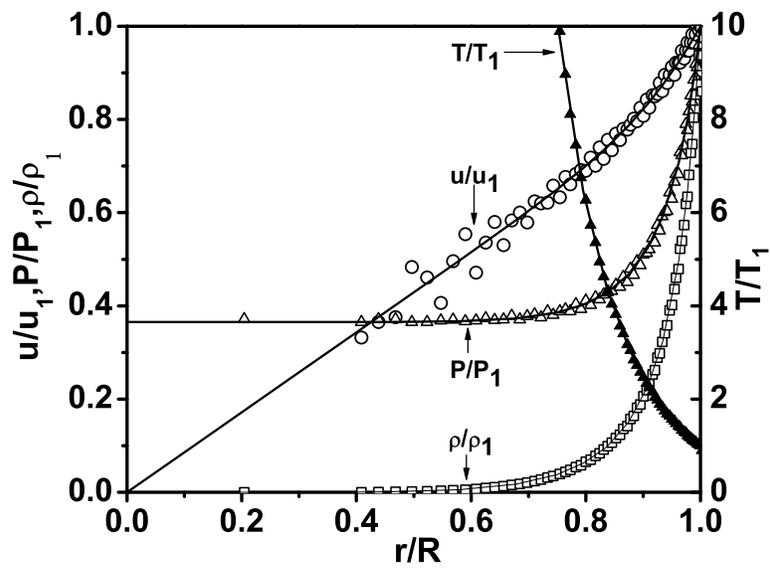}
\end{center}
\caption{\label{figure10}Comparison of the simulation data 
obtained in the pure hydrodynamic case (points) with the 
self similar solutions (lines) for the point explosion problem 
with specific internal energy $E= 10^5\ $Tergs/gm deposited in
 the first two meshes and $\gamma=1.4$ .}
\end{figure}

Also, the distributions of velocity, pressure, density and 
temperature behind the shock front are generated numerically 
using the hydrodynamics code without taking radiation 
interaction into account. Ideal D-T gas of density $\rho_0=1\ gm/cc$
 and $\gamma=1.4$ is filled inside a sphere of 1 cm radius with 
the region divided into 100 radial meshes each of width $0.01\ cm$.
 The initial internal energy per unit mass is chosen as 
$10^5\ Tergs/gm$ for the first two meshes and zero for all 
the other meshes. An initial time step of $10^-6\ \mu s$ is 
chosen and the thermodynamic variables are obtained after a 
time $0.2\ \mu s$. As in the case of the problem of shock 
propagation in aluminium, the total energy equation is 
solved assuming that electrons and ions are at the same 
temperature (the material temperature). In Fig. \ref{figure10}.
 we compare the distribution of the functions 
$P/P_1, u/u_1, \rho/\rho_1\ \mbox{and}\ T/T_1$ with respect to 
$r/R$ obtained exactly by solving the ODEs as explained above 
(solid lines) with the results generated from our code (dots).  
Good agreement between the numerical and theoretical results 
is observed. As is characteristic of a strong explosion, the 
gas density decreases extremely rapidly as we move away from 
the shock front towards the centre as seen from Fig. 
\ref{figure10}. In the vicinity of the front the pressure 
decreases as we move towards the centre by a factor of 2 to
 3 and then remains constant whereas the velocity curve rapidly
 becomes a straight line passing through the origin.The 
temperatures are very high at the centre and decreases 
smoothly at the shock front. As the particles at the centre 
are heated by a strong shock, they have very high entropy 
and hence high temperatures.

\begin{figure}
\includegraphics[width=7.5cm]{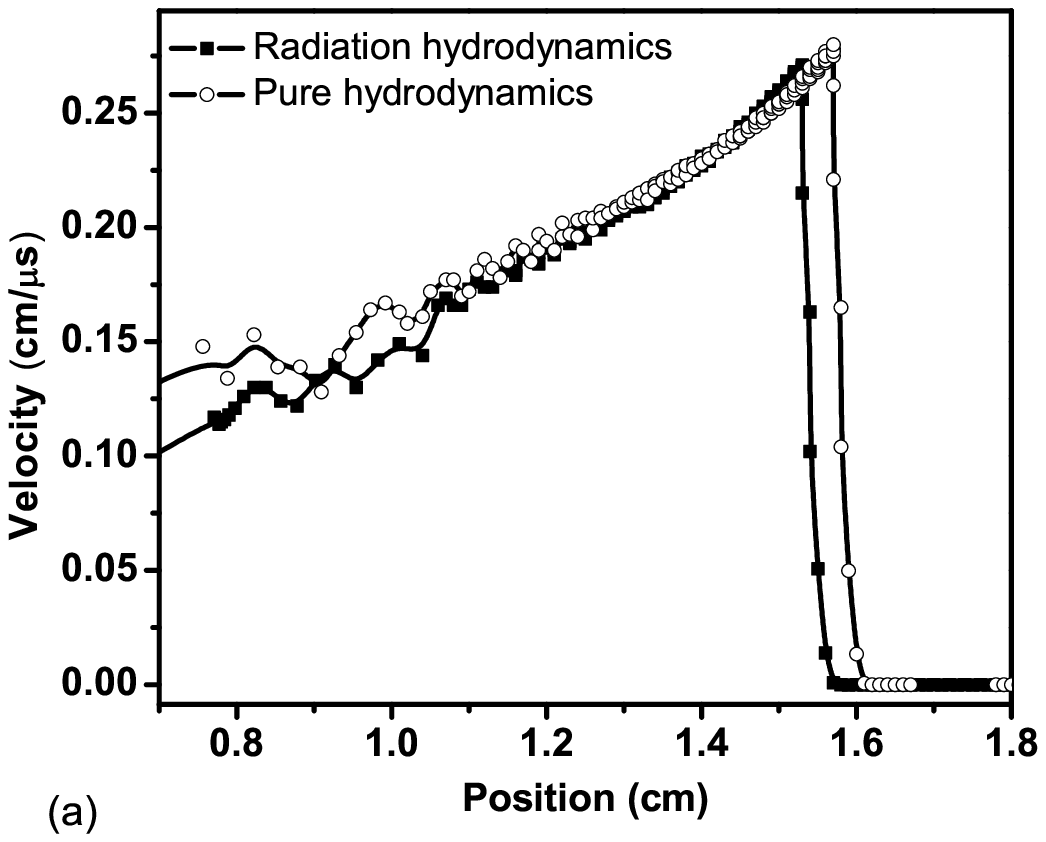}
\includegraphics*[width=7.5cm]{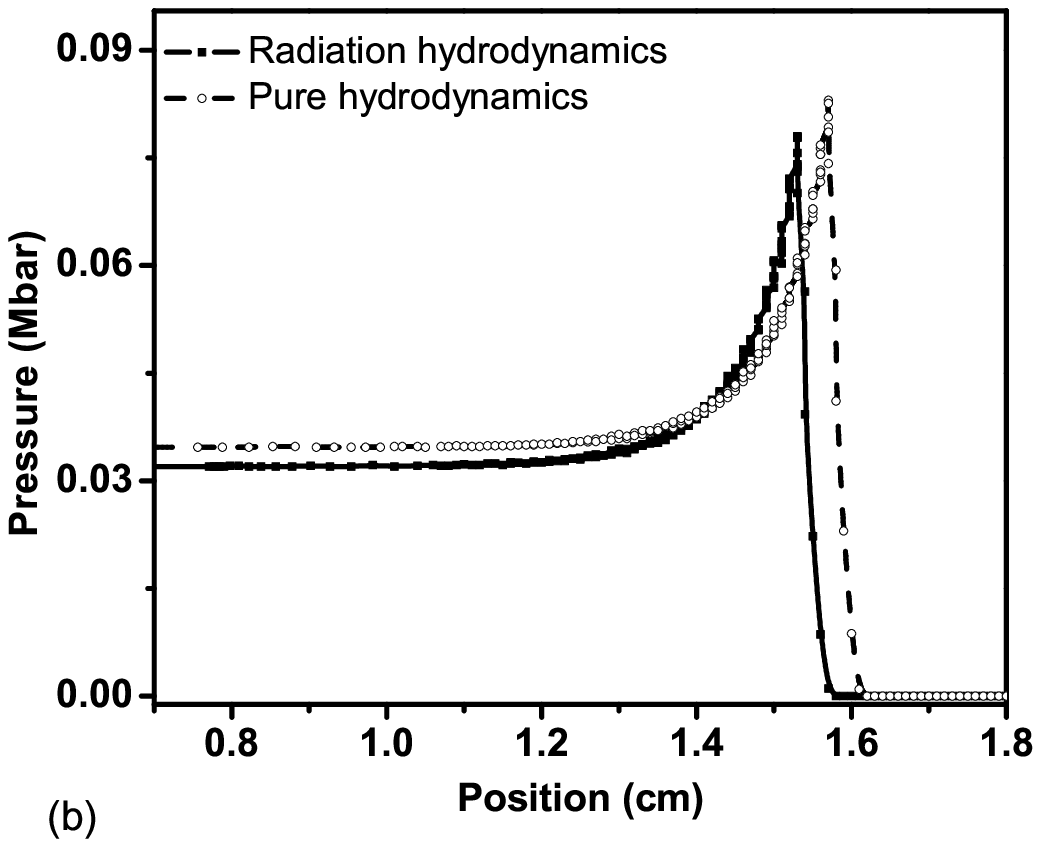}
\includegraphics*[width=7.5cm]{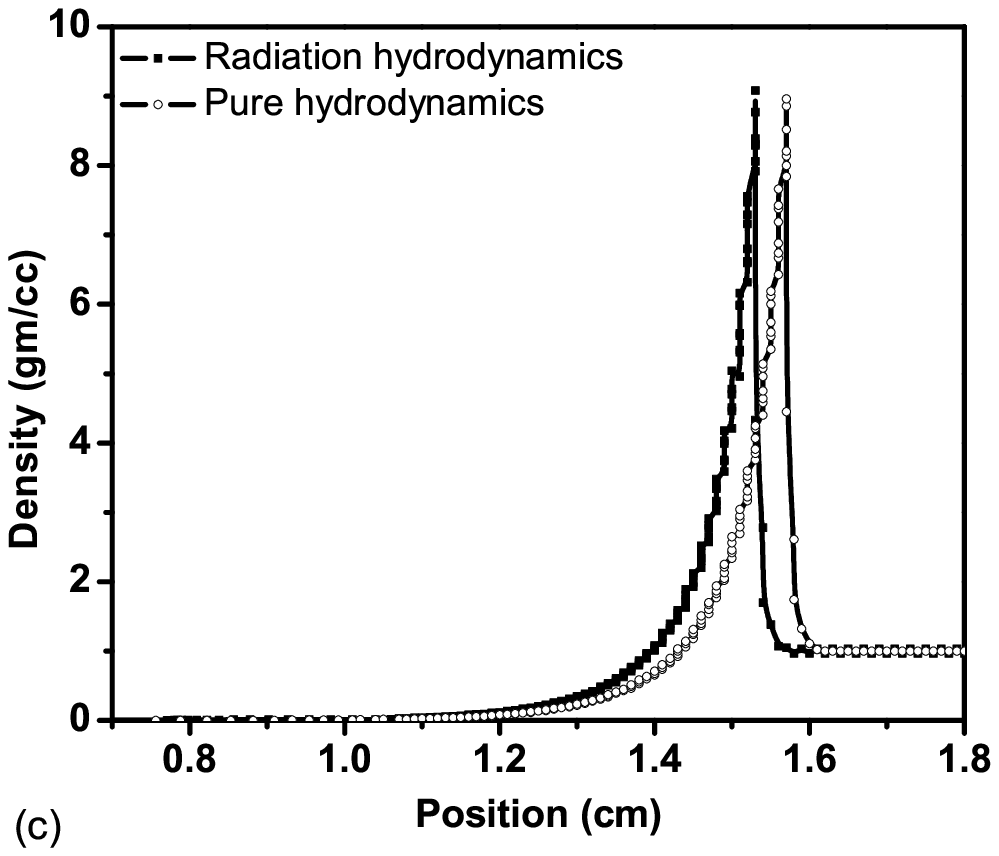}
\includegraphics*[width=7.5cm]{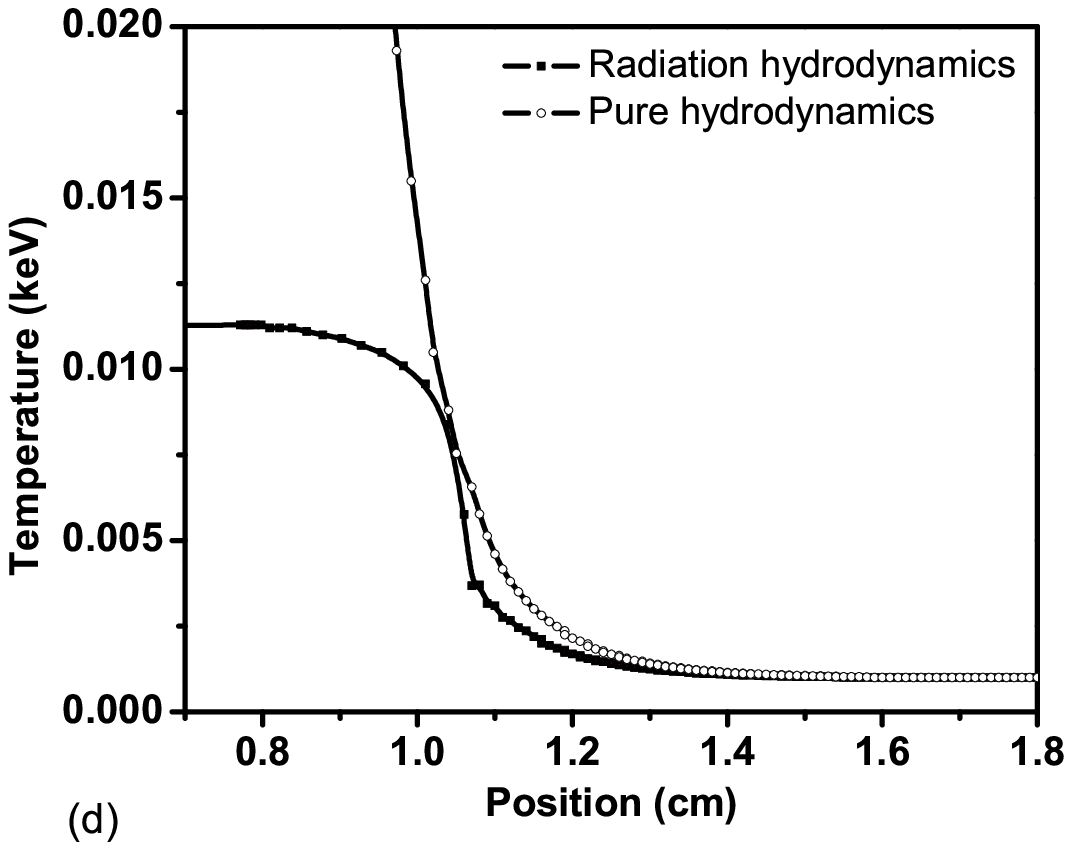}
\caption{\label{figure11}Profiles of the thermodynamic variables 
with and without radiation interaction at $2\ \mu s$ for the point 
explosion problem with specific internal energy $E=10^5\ $Tergs/gm
 deposited in the first two meshes.}
\end{figure}

The radiation interaction effects become prominent only at 
higher temperatures. The Rosseland opacity of a D-T (50:50) 
gas in terms of the gas density and temperature is 
$ \kappa_R = 0.505\ \rho^2\ T^{7/2}$  \cite{Thompson}, 
\cite{Zeldovich}. Fig. \ref{figure11}. shows the profiles 
of velocity, pressure, density and temperature after 
$2\ \mu s$ with an initial specific internal energy of 
$10^5\ Tergs/gm$ deposited in the first two meshes 
(i.e. total energy deposited is 3.351 Tergs 
= $3.351 \times 10^{12} \mbox{ergs}$ ), $\gamma=1.23$
 and 200 meshes each of width $0.01\ cm$. The shock front 
moves faster in the pure hydrodynamic case as no energy is 
lost to the radiation keeping the shock stronger. Also the 
effect of radiation interaction is to reduce the temperature 
at the centre as seen in Fig. \ref{figure11}(d).

\begin{figure}
\includegraphics[width=7.5cm]{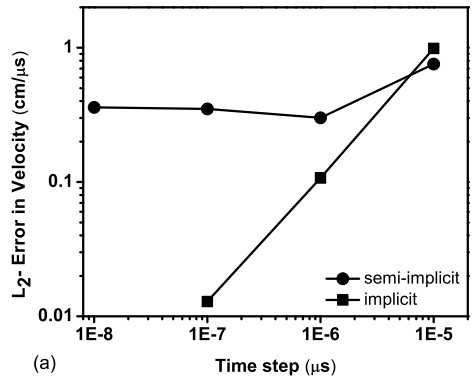}
\includegraphics[width=7.5cm]{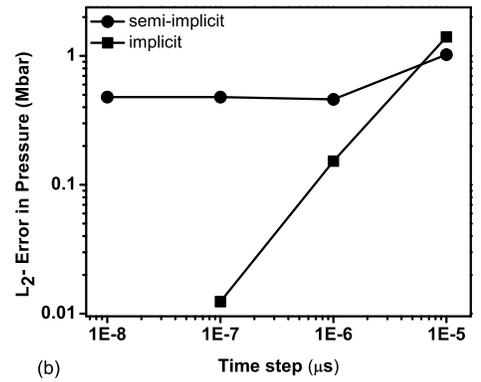}
\includegraphics[width=7.5cm]{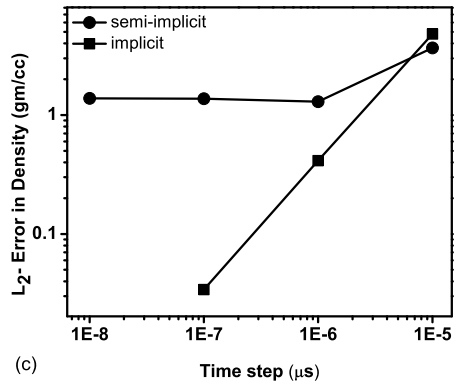}
\includegraphics[width=7.5cm]{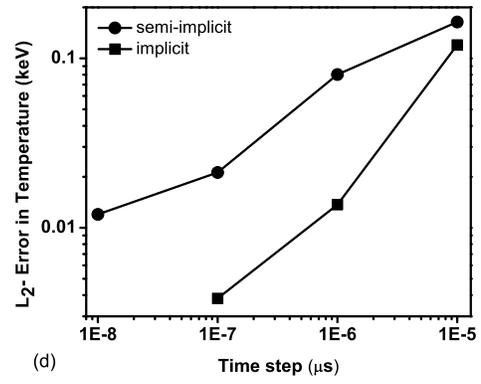}
\caption{\label{figure12}Comparison of the $L_2$-Error Vs. time step
 for the point explosion problem. Implicit scheme converges faster.}
\end{figure}

\begin{figure}
\begin{center}
\includegraphics{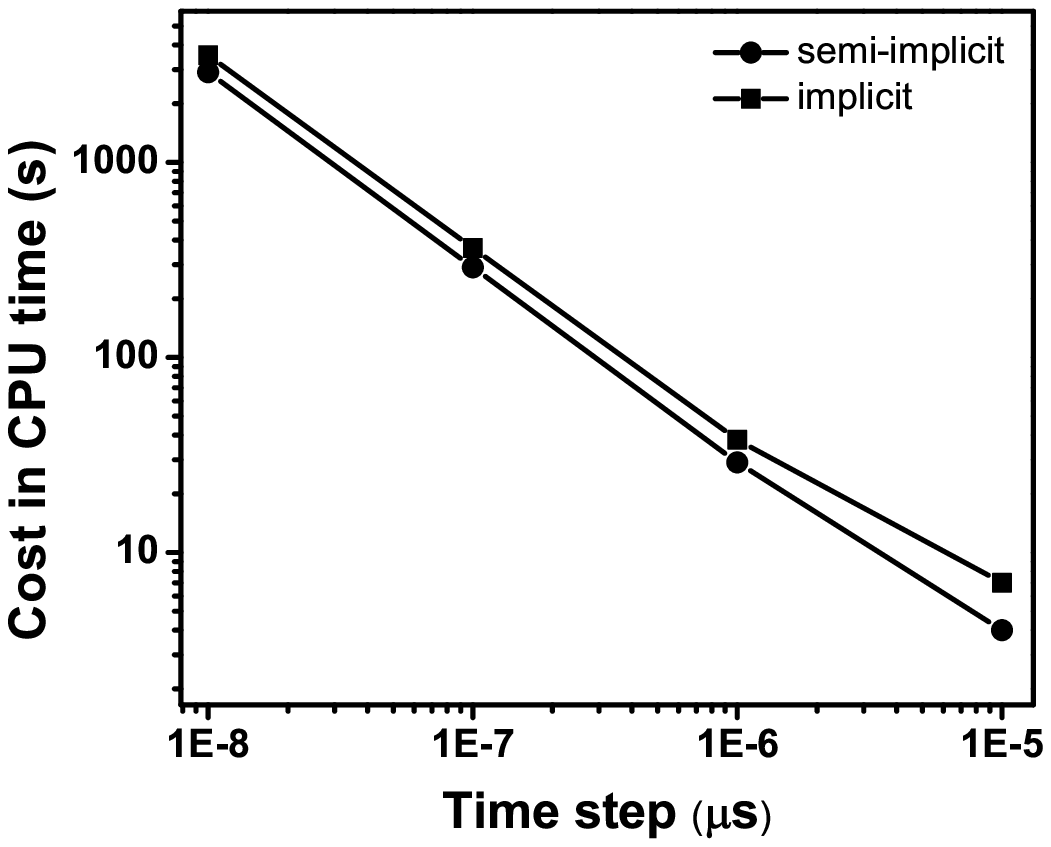}
\end{center}
\caption{\label{figure13}CPU cost for the point explosion problem with 
specific internal energy $E=10^5\ $Tergs/gm deposited in the first two 
meshes for the implicit and semi-implicit schemes taking radiation 
interaction into account.}
\end{figure}

The convergence study for the point explosion problem shows 
that the $L_2$-Error decreases much faster for the implicit method 
compared to the explicit scheme as observed in Fig. \ref{figure12}.
 The exact solution in this case is chosen to be the result from 
a run using the implicit method with a small time step value of 
$10^{-8}\mu s$ and an initial mesh width of $0.01\ cm$. As the 
time step is decreased, the non-linearities are iteratively 
converged for the implicit scheme, whereas for the explicit 
scheme, the spatial discretization error begins to dominate 
thereby preventing any appreciable decrease in the $L_2$-Error 
(as observed in the problem of shock propagation in aluminium 
foil also). In the implicit method, faster convergence is 
attained at the cost of slightly higher CPU time as shown 
in Fig. \ref{figure13}.

\section{Conclusions}\label{conc}
In this paper we have developed and studied the performance of 
fully implicit radiation hydrodynamics scheme as compared to the
 semi-implicit scheme. The time dependent radiation transport 
equation is solved and energy transfer to the medium is 
accounted exactly without invoking approximation methods. 
To validate the code, the results have been verified using 
the problem of shock propagation in Al foil in the planar 
geometry and the point explosion problem in the spherical 
geometry. The simulation results show good agreement with 
the theoretical solutions. The convergence properties show 
that the $L_2$-Error keeps on decreasing on reducing the 
time step value for the implicit scheme. However for the 
semi-implicit scheme, the $L_2$-Error remains fixed below a 
certain time step because of the spatial discretization error 
showing that the implicit scheme is necessary if high accuracy 
is to be obtained. Also the price to be paid for obtaining such 
high accuracies is only slightly higher than the semi-implicit 
scheme in terms of the CPU time. The implicit scheme gives fairly 
accurate result even for large time steps whereas the semi-implicit 
scheme does not work there because of the CFL criterion. 



\newpage

\begin{thebibliography}{00}
\bibitem{Zeldovich}Y.B.Zeldovich and Y.P.Raizer, {\it Physics 
of Shock Waves and High-Temperature Hydrodynamic Phenomena}, 
Vols I and II (Academic Press, New York, 1966)
\bibitem{Mihalas}D.Mihalas and B.W.Mihalas, {\it Foundations of
 Radiation Hydrodynamics} (Oxford Univ.Press, New York,  1984)
\bibitem{Bates}J.W.Bates, et.al., On Consistent Time-Integration
 Methods for Radiation Hydrodynamics in the Equilibrium Diffusion 
Limit: Low-Energy-Density Regime, {\it J.Comput.Phys.} {\bf 167}, 
99 (2001) 
\bibitem{Dai1}W.dai and P.R.Woodward, Numerical simulation for 
radiation hydrodynamics I. Diffusion limit, {\it J.Comput.Phys.} 
{\bf 142}, 182 (1998)
\bibitem{Dai2}W.dai and P.R.Woodward, Numerical simulation for 
radiation hydrodynamics II. Transport limit, {\it J.Comput.Phys.}
 {\bf 157}, 199 (2000)
\bibitem{Knoll}D.A.Knoll, W.J.Rider, G.L.Olson, Nonlinear 
convergence, accuracy, and time step control in nonequilibrium 
radiation diffusion, {\it Journal of Quantitative Spectroscopy 
and Radiative Transfer} {\bf 70}, 25 (2001)
\bibitem{Ober}C.C.Ober and J.N.Shadid, Studies on the accuracy 
of time-integration methods for the radiation-diffusion equations, 
{\it J.Comput.Phys.} {\bf 195}, 743 (2004)
\bibitem{Knoll2}D.A.Knoll, L.Chacon, L.G.Margolin, V.A.Moussean, 
On balanced approximations for time integration of multiple time 
scale systems, {\it J.Comput.Phys.} {\bf 185}, 583 (2003)
\bibitem{Niem}D.de Neim, E.Kuhrt, U.Motschmann, A volume-of-fluid 
method for simulation of compressible axisymmetric multi-material 
flow, {\it Comput. Phys. Commun.} {\bf 176}, 170 (2007)
\bibitem{Von}J. Von Neumann and R.D.Richtmyer, A Method for the 
Numerical Calculation of Hydrodynamic Shocks, {\it J. Appl. Phys.} 
{\bf 21}, 232 (1950)
\bibitem{Huba}Huba J D. 2006 {\it NRL Plasma Formulary} 
(Washington: Naval Research Lab.) 35
\bibitem{Larsen}E.Larsen, A Grey Transport Acceleration 
Method for Time-Dependent Radiative Transfer Problems, 
{\it J.Comput.Phys.} {\bf 78}, 459 (1988)
\bibitem{Lewis}E.E.Lewis and W.F.Miller, Jr., {\it Computational 
Methods of Neutron Transport}, (John Wiley and Sons, New York, 1984)
\bibitem{Bar}P.Barbucci and F.Di Pasquantonio, Exponential 
Supplementary Equations for $S_N$ Methods: The One-Dimensional 
Case, {\it Nucl. Sci. Eng.} {\bf 63}, 179 (1977)
\bibitem{Wilkins}M.L.Wilkins, {\it Computer Simulation 
of Dynamic Phenomena} (Springer-Verlag, Berlin, Heidelberg, 
New York, 1999) ISBN 3-540-63070-8 
\bibitem{Richt}R.D.Richtmyer and K.W.Morton, {\it Difference 
Methods for Initial-Value Problems}, Second Edition (Interscience 
Publishers, New York, 1967) 
\bibitem{Kauffman}R.L.Kauffman, et.al, High Temperatures in 
Inertial Confinement Fusion Radiation Cavities heated with 
$0.35 \mu m$ light, {\it Phys. Rev. Lett.} {\bf 73}, 2320 (1994)
\bibitem{Basko}M.Basko, An improved version of the view factor 
method for simulating inertial confinement fusion hohlraums, 
{\it Phys. Plasmas} {\bf 3}, 4148 (1996) 
\bibitem{Sedov}L.I.Sedov, {\it Similarity and Dimensional 
Methods in Mechanics} (Gostekhizdat, Moscow, 4 th edition 
1957. English transl. (M.Holt, ed.), Academic Press, New 
York, 1959)
\bibitem{Thompson}S.L.Thompson and H.S.Lauson, Improvements 
in the Chart D Radiation-Hydrodynamic Code III: Revised Analytic 
Equation of State, Sandia Laboratories, {\bf SC-RR-71 0714} (1972)




\end{thebibliography}
\end{document}